\documentclass[prd,showpacs,floatfix,twocolumn,,amsmath,amssymb,floatfix]{revtex4}
\usepackage{graphicx,color,dcolumn,booktabs,bm}
\usepackage{longtable,lscape}
\usepackage{txfonts}
\usepackage{overpic}
\usepackage{amssymb}
\usepackage{indentfirst}
\usepackage{feynmf}   
\usepackage{slashed}  
\usepackage{cases}
\usepackage{color}
\usepackage{multirow}
\usepackage{graphicx,color,dcolumn,booktabs,bm}

\graphicspath{{Figures/}} %

\graphicspath{{Figures/}} %

\usepackage[colorlinks, citecolor=blue,anchorcolor=red,menucolor=red, linkcolor=red,filecolor=red,runcolor=red,urlcolor=blue,frenchlinks=red]{hyperref}

\allowdisplaybreaks

\begin{document}

\title{$Z_{cs}(3985)^-$: a strange hidden-charm tetraquark resonance or not?}

\author{Rui Chen$^{1,2}$}\email{chen$_$rui@pku.edu.cn}
\author{Qi Huang$^{3}$}

\affiliation{
$^1$Center of High Energy Physics, Peking University, Beijing
100871, China\\
$^2$School of Physics and State Key Laboratory of Nuclear Physics and Technology, Peking University, Beijing 100871, China\\
$^3$School of Physical Sciences, University of Chinese Academy of Sciences, Beijing 100049, China}
\date{\today}

\begin{abstract}
  Inspired by the newly $Z_{cs}(3985)^-$ reported by the BESIII Collaboration in the $K^+$ recoil-mass spectrum of the of $e^+e^-\to (D^{*0}D_s^-/D^0D_s^{*-})K^+$ processes, we perform a dynamical study on the $D^{(*)0}D_s^{*-}$ interactions by adopting a one-boson-exchange model and considering the coupled channel effect. After producing the phase shifts for all the discussed channels, our results exclude the newly $Z_{cs}(3985)^-$ as a $D^{*0}D_s^{-}/D^{0}D_s^{*-}/D^{*0}D_s^{*-}$ resonance with $I(J^P)=1/2(1^+, 0^-, 1^-, 2^-)$.
\end{abstract}


\maketitle

\section{introduction}



Since the BESIII Collaboration reported a discovery of $Z_{c}(3900)$ in $e^+e^-\to\pi^+\pi^-J/\psi$ process in 2013 \cite{Ablikim:2013mio}, in the past decade, a series of charged and neutral $Z_c$ states were observed \cite{Liu:2013dau,Xiao:2013iha,Ablikim:2013wzq,Ablikim:2013xfr,Ablikim:2013emm,Ablikim:2017oaf}. The discoveries of these $Z_c$ states suddenly caught much attention as their novel properties. For example, $Z_c(3900)$ was observed in the $J/\psi \pi^\pm$ invariant mass spectrum of the $e^+e^-\to\pi^+\pi^-J/\psi$ process at $\sqrt{s}= 4.26$ GeV, due to its charged property and decay final states, $Z_c(3900)$ should have exotic quark configurations, e.g., it is different with the conventional mesons $(q\bar{q})$ and baryons $(qqq)$. Thus, these $Z_c$ states bring great challenges for the conventional quark model and understanding of strong interactions.

Very recently, the BESIII Collaboration did an analysis on the processes of $e^+e^-\to (D^{*0}D_s^-/D^0D_s^{*-})K^+$ and observed a structure $Z_{cs}^-(3985)$ in the $K^+$ recoil-mass spectrum when $\sqrt{s}=4.681$ GeV \cite{Ablikim:2020hsk}, whose pole position is
\begin{eqnarray*}
  M &=& 3982.5_{-2.6}^{+1.8}\pm2.1 \text{MeV}, \\
  \Gamma &=& 12.8_{-4.4}^{+5.3}\pm3.0 \text{MeV}.
\end{eqnarray*}
The observation of this $Z_{cs}(3985)$ obtains widely discussions \cite{Wan:1,Yang:2,Meng:3,Wang:4}. Similar to our above description on $Z_c(3900)$, it is easy to get that it may have four different valence quark components $[c\bar{c}s\bar{u}]$. Thus, it is the first candidate of the charged hidden-charm tetraquark state with strangeness. Actually, as mentioned in \cite{Ablikim:2020hsk}, it is a partner of currently existing $Z_c(3885)$.

\begin{figure}[!htbp]
\center
\includegraphics[width=3.4in]{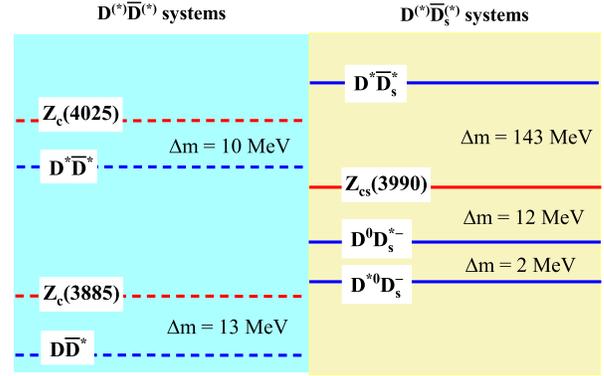}\\
\caption{(color online) A mass comparison between $Z_c$ ($Z_{cs}$) states and thresholds of a charmed meson and an anti-charmed (charm-strange) meson.}\label{mass}
\end{figure}

As shown in Fig. \ref{mass}, the $Z_{cs}(3985)$ is a little above the $D^{*0}D_s^-$ and $D^0D_s^{*-}$ thresholds (right panel), compared to the positions of the $Z_c(3885)$ and $Z_c(4025)$ (left panel), we can easily see its similarity with the $Z_c(3885)$ and $Z_c(4025)$. Since the $Z_c(3885)$ and $Z_c(4020)$ have often been assigned as hidden-charm hadronic molecules or resonances \cite{Chen:2016qju,Liu:2013waa,Guo:2017jvc,Brambilla:2019esw}, one may propose whether the newly $Z_{cs}(3985)$ can be a strange hidden-charm tetraquark resonance. In Ref. \cite{Ebert:2008kb}, the mass of the $cq\bar{c}\bar{s}$ was predicted as 3982 MeV within the relativistic diquark-antidiquark picture. In this work, we study the $D^{(*)}\bar{D}_s^{(*)}$ interactions by using the one-boson-exchange (OBE) model to examine this possibility.

In general, the resonances can be divided into two types, e.g., shape-type and Feshbach-type resonances. As we known, the potential barriers can be beneficial to generate these two types of resonances. In particular, we need to mention that the coupled channel effect plays a very important role in the generation of Feshbach-type resonances, where the mass gaps between the discussed channels contribute to the barriers.

In the following, we will consider the coupled-channel effect and $S-D$ wave mixing effect, which can provide mass gap barriers and centrifugal potential barriers, respectively. Since the spin-parity of $Z_{cs}$ isn't fixed yet, the different spin-parities $I(J^P)=1/2(1^+, 0^-, 1^-, 2^-)$ for the coupled $D^{(*)}\bar{D}_s^{(*)}$ systems will be discussed.


This paper is organized as follows. After this introduction, we present the $D^{(*)}\bar{D}_s^{(*)}$ interactions by using the OBE model in Sec.~\ref{sec2}. The corresponding numerical results are given in Sec.~\ref{sec3}. The paper ends with a summary in Sec. \ref{sec4}.

\section{Interactions}\label{sec2}

For the coupled $D^{*0}{D}_s^{-}/D^{0}D_s^{*-}/D^{*0}D_s^{*-}$ system, in the OBE model, there only exists contributions of $\sigma$ and $\eta$ exchanges. Here, we need to point out that the $K/K^*-$exchanges are forbidden in the cross processes $D^{*}\bar{D}_s\to D_s^{(*)}\bar{D}^{(*)}$. Since the descriptions of interactions between charmed mesons and Goldstone bosons are needed, we adopt the effective Lagrangians approach. The effective Lagrangians relevant to the charmed hadrons and light mesons are constructed as
\begin{eqnarray}
\mathcal{L} &=&
g_{\sigma} \left\langle H^{(Q)}_a\sigma\overline{H}^{(Q)}_a\right\rangle
   +ig\left\langle H^{(Q)}_b\gamma_{\mu}{A}_{ba}^{\mu}\gamma_5 \overline{H}^{(Q)}_a\right\rangle\nonumber\\
 &&+g_{\sigma} \left\langle \overline{H}^{(\bar{Q})}_a\sigma H^{(\bar{Q})}_a\right\rangle
+ig\left\langle \overline{H}_a^{(\bar{Q})}\gamma_{\mu}{A}_{ab}^{\mu}\gamma_5H_b^{(\bar{Q})}\right\rangle,\,\label{lag1}
\end{eqnarray}
which are based on the heavy-quark limit and chiral symmetry \cite{Yan:1992gz,Wise:1992hn,Burdman:1992gh,Casalbuoni:1996pg,Falk:1992cx,Liu:2011xc}. Here, $A_{\mu}=(\xi^{\dag}\partial_{\mu}\xi-\xi\partial_{\mu}\xi^{\dag})/2=i
\partial_{\mu}{\mathbb{P}}/f_{\pi}+\ldots$ is the axial current, with the pion decay constant $f_{\pi}= 132$ MeV and $\xi = \text{exp}(i\mathbb{P}/f_{\pi})$. In addition, $g=0.59$ is fixed by fitting the decay width of $D^{*+}$ \cite{Tanabashi:2018oca}, and $g_s=0.76$ is given in Refs. \cite{Falk:1992cx}. In Eq. (\ref{lag1}), we define two multiplet fields $H^{({Q})}$ and $H^{(\bar{Q})}$, which are expressed as a linear combination of a pseudoscalar charmed (anti-charmed) meson and a vector charmed (anti-charmed) meson, respectively,
\begin{eqnarray}
H_a^{(Q)} &=& \frac{1+\rlap\slash v}{2}\left[\mathcal{P}_a^{*(Q)\mu}\gamma_{\mu}-\mathcal{P}_a^{(Q)}\gamma_5\right],\nonumber\\
H_a^{(\bar{Q})} &=& [\widetilde{\mathcal{P}}_a^{*\mu}\gamma_{\mu}-\widetilde{\mathcal{P}}_a\gamma_5]\frac{1-\rlap\slash v}{2}.\label{lag2}
\end{eqnarray}
Here, $\bm{v}=(1, \bm{0})$, $\mathcal{P}^{(*)T}=(D^{(*)+}, D^{(*)0}, D_s^{(*)+})$, and $\widetilde{\mathcal{P}}^{(*)T}=(D^{(*)-}, \bar{D}^{(*)0}, D_s^{(*)-})$. The pseudoscalar matrix $\mathbb{P}$ reads as $\mathbb{P}=\text{diag}(\frac{\eta}{\sqrt{6}}, \frac{\eta}{\sqrt{6}}, -\frac{2\eta}{\sqrt{6}})$.

After expanding Eqs. (\ref{lag1})-(\ref{lag2}), we obtain the concrete effective Lagrangians as follows,
\begin{eqnarray}
\mathcal{L}_{\sigma} &=& -2g_{\sigma}\mathcal{P}_a\mathcal{P}_a^{\dag}\sigma-2g_{\sigma}\bar{\mathcal{P}}_a^{\dag}\bar{\mathcal{P}}_a\sigma\nonumber\\
  &&+2g_{\sigma}\mathcal{P}_a^*\cdot \mathcal{P}_a^{*\dag}\sigma
            +2g_{\sigma}\bar{\mathcal{P}}_a^{*\dag}\cdot\bar{\mathcal{P}}_a^{*}\sigma,\nonumber\\
\mathcal{L}_{\mathbb{P}} &=&
 -\frac{2g}{f_{\pi}}\left(\mathcal{P}_{b}^{*\mu}\mathcal{P}_{a}^{\dag}
 +\mathcal{P}_{b}\mathcal{P}_{a}^{*\mu\dag}\right)\partial_{\mu}\mathbb{P}_{ba}\nonumber\\
 &&+\frac{2g}{f_{\pi}}\left(\bar{\mathcal{P}}_{a}^{*\mu\dag}\bar{\mathcal{P}}_{b}
 +\bar{\mathcal{P}}_{a}^{\dag}\bar{\mathcal{P}}_{b}^{*\mu}\right)
            \partial_{\mu}\mathbb{P}_{ab}\nonumber\\
           &&-i\frac{2g}{f_{\pi}}v^{\beta}\varepsilon_{\beta\mu\alpha\nu}
           \mathcal{P}_{b}^{*\mu}\mathcal{P}_{a}^{*\nu\dag}\partial^{\alpha}\mathbb{P}_{ba}\nonumber\\
           &&+i\frac{2g}{f_{\pi}}v^{\beta}\varepsilon_{\beta\mu\alpha\nu}
           \bar{\mathcal{P}}_{a}^{*\mu\dag}\bar{\mathcal{P}}_{b}^{*\nu}\partial^{\alpha}\mathbb{P}_{ab}.
\end{eqnarray}

With these effective Lagrangians, we can easily deduce the corresponding OBE effective potentials, which are related to the scattering amplitudes by using a Breit approximation in a momentum space, i.e.,
\begin{eqnarray}\label{breit}
\mathcal{V}_{E}^{h_1h_2\to h_3h_4}(\bm{q}) &=&
          -\frac{\mathcal{M}(h_1h_2\to h_3h_4)}
          {\sqrt{\prod_i2M_i\prod_f2M_f}}.
\end{eqnarray}
Here, $M_i$ and $M_f$ are the masses of the initial states ($h_1$, $h_2$) and final states ($h_3$, $h_4$), respectively. $\mathcal{M}(h_1h_2\to h_3h_4)$ is the scattering amplitude for the $h_1h_2\to h_3h_4$ process. After performing a Fourier transformation, we can obtain the effective potential in the coordinate space $\mathcal{V}(\bm{r})$, i.e.,
\begin{eqnarray}
\mathcal{V}_{E}^{h_1h_2\to h_3h_4}(\bm{r}) =
          \int\frac{d^3\bm{q}}{(2\pi)^3}e^{i\bm{q}\cdot\bm{r}}
          \mathcal{V}_{E}^{h_1h_2\to h_3h_4}(\bm{q})\mathcal{F}^2(q^2,m_E^2),\nonumber
\end{eqnarray}
where we introduce a monopole form factor $\mathcal{F}(q^2,m_E^2)=(\Lambda^2-m_E^2)/(\Lambda^2-q^2)$ at every interactive vertex, which compensates the off-shell effect of the exchanged bosons. Here, $\Lambda$, $m_E$, and $q$ denote the cutoff, mass, and four-momentum of the exchanged meson, respectively.

Finally, the OBE effective potentials for the coupled $D^{*0}{D}_s^{-}/D^{0}D_s^{*-}/D^{*0}D_s^{*-}$ systems are
\begin{widetext}
\begin{eqnarray}
\mathcal{V}_{\text{OBE}} &=& \left(\begin{array}{ccl}
  \langle D^{*0}{D}_s^{-}|\mathcal{V}_{\sigma}|D^{*0}{D}_s^{-}\rangle
        &\langle D^{*0}{D}_s^{-}|\mathcal{V}_{\eta}|D^{0}D_s^{*-}\rangle
            &\langle D^{*0}{D}_s^{-}|\mathcal{V}_{\eta}|D^{*0}{D}_s^{*-}\rangle\\
  \langle D^{0}{D}_s^{*-}|\mathcal{V}_{\eta}|D^{*0}{D}_s^{-}\rangle
        &\langle D^{0}{D}_s^{*-}|\mathcal{V}_{\sigma}|D^{0}{D}_s^{*-}\rangle
            &\langle D^{0}{D}_s^{*-}|\mathcal{V}_{\eta}|D^{*0}{D}_s^{*-}\rangle\\
  \langle D^{*0}{D}_s^{*-}|\mathcal{V}_{\eta}|D^{*0}{D}_s^{-}\rangle
        &\langle D^{0}{D}_s^{*-}|\mathcal{V}_{\eta}|D^{0}{D}_s^{*-}\rangle
            &\langle D^{*0}{D}_s^{*-}|\mathcal{V}_{\sigma}+\mathcal{V}_{\eta}|D^{*0}{D}_s^{*-}\rangle\end{array}\right).,\label{potential}
\end{eqnarray}
\end{widetext}
where the subpotentials are
\begin{eqnarray}
\mathcal{V}_{\sigma}^{D^{*0}D_{s}^-\to D^{*0}D_s^-} &=& -g_s^2\bm{\epsilon}_1\cdot\bm{\epsilon}_3^{\dag}Y(\Lambda,m_{\sigma},r),\label{po1}\\
\mathcal{V}_{\sigma}^{D^{0}D_{s}^{*-}\to D^{0}D_s^{*-}} &=& -g_s^2\bm{\epsilon}_2\cdot\bm{\epsilon}_4^{\dag}Y(\Lambda,m_{\sigma},r),\\
\mathcal{V}_{\sigma}^{D^{*0}D_{s}^{*-}\to D^{*0}D_s^{*-}} &=& -g_s^2\bm{\epsilon}_1\cdot\bm{\epsilon}_3^{\dag}\bm{\epsilon}_2\cdot\bm{\epsilon}_4^{\dag}Y(\Lambda,m_{\sigma},r),\\
\mathcal{V}_{\eta}^{D^{*0}D_{s}^-\to D^{0}D_s^{*-}} &=& \frac{1}{9}\frac{g^2}{f_{\pi}^2}\left[\bm{\epsilon}_1\cdot\bm{\epsilon}_4^{\dag}Z(\Lambda_1,m_{\eta1},r)\right.\nonumber\\ &&\left.+S(\hat{r},\bm{\epsilon}_1,\bm{\epsilon}_4^{\dag})T(\Lambda_1,m_{\eta1},r)\right],\\
\mathcal{V}_{\eta}^{D^{*0}D_{s}^-\to D^{*0}D_s^{*-}} &=& \frac{1}{9}\frac{g^2}{f_{\pi}^2}\left[i(\bm{\epsilon}_1\times\bm{\epsilon}_3^{\dag})\cdot\bm{\epsilon}_4^{\dag}
Z(\Lambda_2,m_{\eta2},r)\right.\nonumber\\
&&\left.+S(\hat{r},i(\bm{\epsilon}_1\times\bm{\epsilon}_3^{\dag}),\bm{\epsilon}_4^{\dag})T(\Lambda_2,m_{\eta2},r)\right],\,\\
\mathcal{V}_{\eta}^{D^{0}D_{s}^{*-}\to D^{*0}D_s^{*-}} &=& \frac{1}{9}\frac{g^2}{f_{\pi}^2}\left[i(\bm{\epsilon}_2\times\bm{\epsilon}_3^{\dag})\cdot\bm{\epsilon}_4^{\dag}
Z(\Lambda_3,m_{\eta3},r)\right.\nonumber\\
&&\left.+S(\hat{r},i(\bm{\epsilon}_2\times\bm{\epsilon}_3^{\dag}),\bm{\epsilon}_4^{\dag})
T(\Lambda_3,m_{\eta3},r)\right],\\
\mathcal{V}_{\eta}^{D^{*0}D_{s}^{*-}\to D^{*0}D_s^{*-}} &=& \frac{1}{9}\frac{g^2}{f_{\pi}^2}\left[(\bm{\epsilon}_1\times\bm{\epsilon}_3^{\dag})
\cdot(\bm{\epsilon}_2\times\bm{\epsilon}_4^{\dag})Z(\Lambda,m_{\eta},r)\right.\nonumber\\ &&\left.+S(\hat{r},\bm{\epsilon}_1\times\bm{\epsilon}_3^{\dag},\bm{\epsilon}_2\times\bm{\epsilon}_4^{\dag})
T(\Lambda,m_{\eta},r)\right],\quad\label{po2}
\end{eqnarray}
with
\begin{eqnarray*}
&&Y(\Lambda,m,r) = \frac{1}{4\pi r}(e^{-mr}-e^{-\Lambda r})-\frac{\Lambda^2-m^2}{8 \pi\Lambda}e^{-\Lambda r},\\
&&T(\Lambda,m,r)=r\frac{\partial}{\partial r}\frac{1}{r}\frac{\partial}{\partial r}Y(\Lambda,m,r),\\
&&Z(\Lambda,m,r)=\nabla^2Y(\Lambda,m,r)=\frac{1}{r^2}\frac{\partial}{\partial r}r^2\frac{\partial}{\partial r}Y(\Lambda,m,r).
\end{eqnarray*}

In Eqs. (\ref{po1})-(\ref{po2}), for conciseness, we introduce many variables, and their definitions are
\begin{eqnarray*}
&&\Lambda_1^2 = \Lambda^2-q_1^2,\quad\quad\quad\quad
m_{\eta1}^2=m_{\eta}^2-q_1^2,\\
&&q_1 = \frac{(m_{D^*}^2+m_{D_s^*}^2)-(m_{D_s}^2+m_{D}^2)}{2(m_{D}+m_{D_s^{*}})},\\
&&\Lambda_2^2 = \Lambda^2-q_2^2,\quad\quad\quad\quad
m_{\eta2}^2=m_{\eta}^2-q_2^2,\\
&&\Lambda_3^2 = \Lambda^2-q_3^2,\quad\quad\quad\quad
m_{\eta3}^2=m_{\eta}^2-q_3^2,\\
&&q_2 = \frac{m_{D_s^*}^2-m_{D_s}^2}{2(m_{D^*}+m_{D_s^{*}})},\quad\,\,
q_3 = \frac{m_{D}^2-m_{D^*}^2}{2(m_{D^*}+m_{D_s^{*}})}.
\end{eqnarray*}
Here, we also define several spin-spin interaction and tensor force operators, in the following numerical calculations, they should be sandwiched by the spin-orbit wave functions for the coupled $D^{*0}{D}_s^{-}/D^{0}D_s^{*-}/D^{*0}D_s^{*-}$ systems, i.e., $\langle{}^{2S^\prime+1}L^{\prime}_{J^{\prime}}|\hat{\mathcal{O}}|{}^{2S+1}L_{J}\rangle$. For the $J^P=1^+$ case, the spin-orbit wave functions $|{}^{2S+1}L_J\rangle$ are $D^{*0}{D}_s^{-}|{}^3S_1,{}^3D_1\rangle$, $D^{0}D_s^{*-}|{}^3S_1,{}^3D_1\rangle$, and $D^{*0}D_s^{*-}|{}^3S_1,{}^3D_1\rangle$. Therefore, the matrix elements for all the operators are
\begin{eqnarray}
&&\left\{\begin{array}{c}
\bm{\epsilon}_1\cdot\bm{\epsilon}_3^{\dag}\\
\bm{\epsilon}_1\cdot\bm{\epsilon}_4^{\dag}\\
\bm{\epsilon}_1\cdot\bm{\epsilon}_3^{\dag}\bm{\epsilon}_2\cdot\bm{\epsilon}_4^{\dag}\\
\frac{i}{\sqrt{2}}(\bm{\epsilon}_2\times\bm{\epsilon}_3^{\dag})\cdot\bm{\epsilon}_4^{\dag}\\
(\bm{\epsilon}_1\times\bm{\epsilon}_3^{\dag})\cdot(\bm{\epsilon}_2\times\bm{\epsilon}_4^{\dag})\end{array}
\right\} \to \left(\mathcal{I}\right),\\
&&\left\{\begin{array}{c}
S(\hat{r},\bm{\epsilon}_1,\bm{\epsilon}_4^{\dag})\\
-\sqrt{2}S(\hat{r},i\bm{\epsilon}_2\times\bm{\epsilon}_3^{\dag},\bm{\epsilon}_4^{\dag})\\
S(\hat{r},\bm{\epsilon}_1\times\bm{\epsilon}_3^{\dag},\bm{\epsilon}_2\times\bm{\epsilon}_4^{\dag})\end{array}
\right\}\to \left(\begin{array}{cc}0    &-\sqrt{2}\\  -\sqrt{2}   &1\end{array}\right),
\end{eqnarray}
where $\mathcal{I}$ stands for unit matrix.

\section{Numerical results}\label{sec3}

In the above OBE effective potentials, there is a cutoff parameter needed to be fixed. According to the quantitative description of the deuteron properties and the $NN$ scattering data (see review in Ref. \cite{Machleidt:1987hj}), the value of cutoff $\Lambda$ is taken around $1$ to $2$ GeV. Here, we will adopt this empirical value to give our conclusion.

\begin{figure}[!htbp]
\center
\includegraphics[width=2.4in]{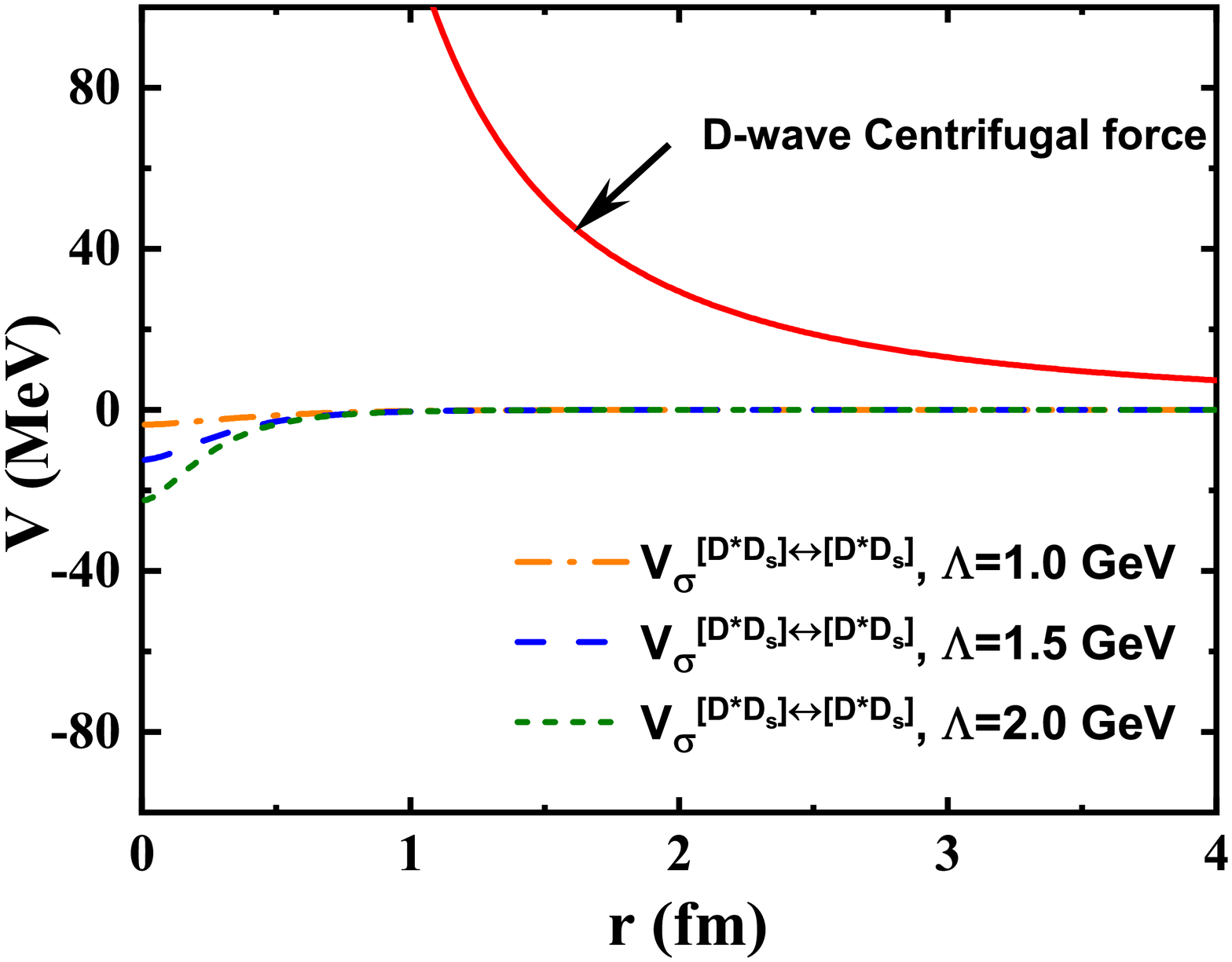}\\
\caption{(color online) OBE effective potential for the $S$ wave $D^{*0}{D}_s^{-}$ system with $\Lambda=1.0, 1.5, 2.0$ GeV and its centrifugal force $l(l+1)/2mr^2$ in $D$ wave $l=2$.}\label{swave}
\end{figure}

In Fig. \ref{swave}, we present the OBE effective potentials for the $S$ wave $D^{*0}{D}_s^{-}$ system with several typical cutoff values. Here, we find that the interaction from the $\sigma$ exchange is weak attractive, and it becomes stronger with the increasing of cutoff $\Lambda$. It is obvious that this interaction isn't stronger enough to form a bound state in the current cutoff choices.

As above mentioned, a repulsive barrier plays an important role in generating a resonance. However, as we can see from Fig. \ref{swave}, there doesn't exist any barrier in the attractive interaction from the $\sigma$ exchange, thus, the $S$ wave $D^{*0}{D}_s^{-}$ cannot bind as a scattering state. So, we further introduce the $S-D$ wave mixing since it may be helpful to generate a $D^{*0}{D}_s^{-}$ scattering state with $I(J^P)=1/2(1^+)$ because the centrifugal force $l(l+1)/2mr^2$ from the $D$ wave $l=2$ can provide a repulsive potential barrier as shown in Fig. \ref{swave}. However, comparing to the weak attractive $\sigma$ exchange interaction, the repulsive centrifugal force is too much stronger. Therefore, we can give an ambitious estimation that there cannot exist a $D^{*0}{D}_s^{-}$ scattering state with $I(J^P)=1/2(1^+)$ when we only consider the single $D^{*0}{D}_s^{-}$ systems.

\begin{figure*}[!htbp]
\center
\includegraphics[width=2.3in]{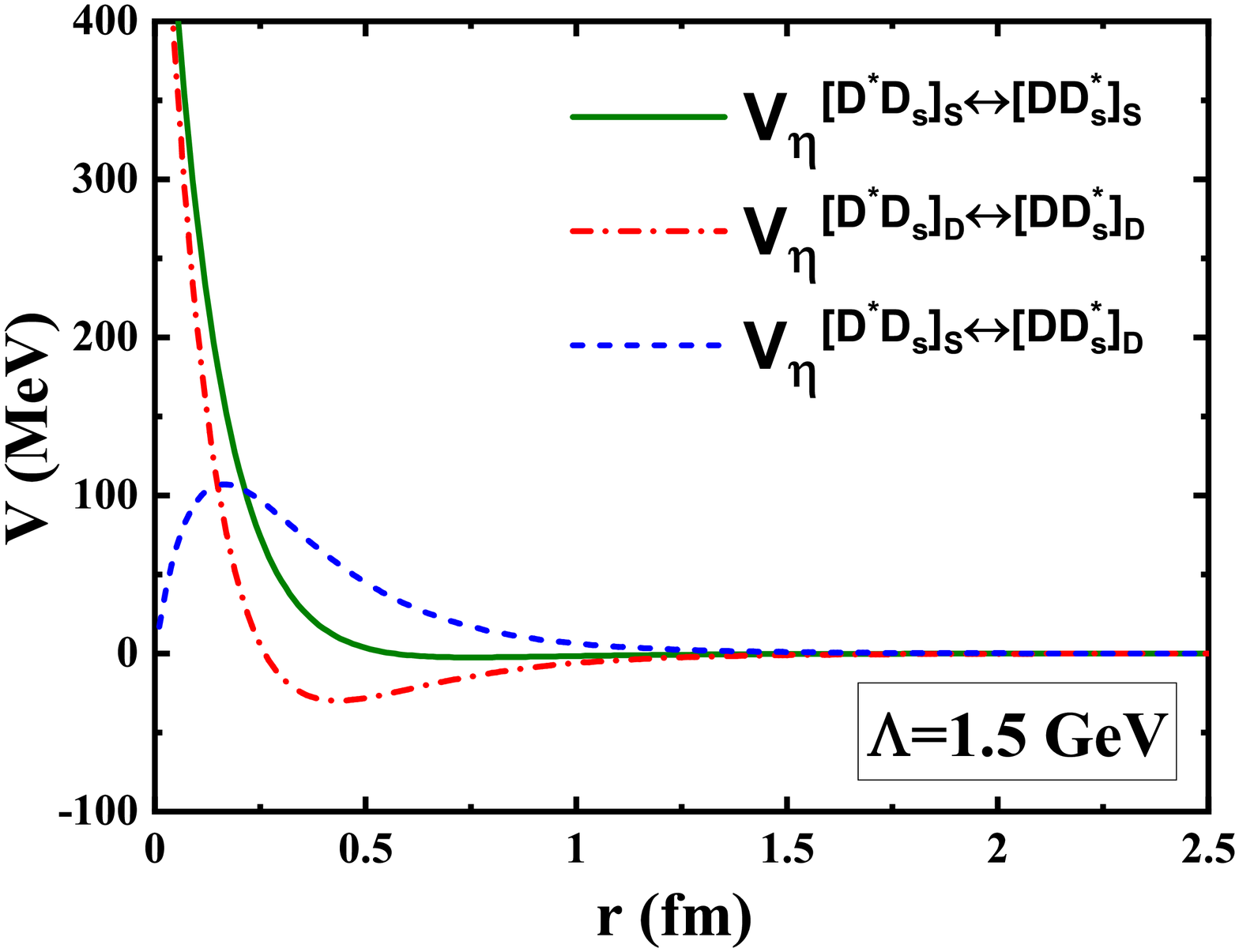}
\includegraphics[width=2.3in]{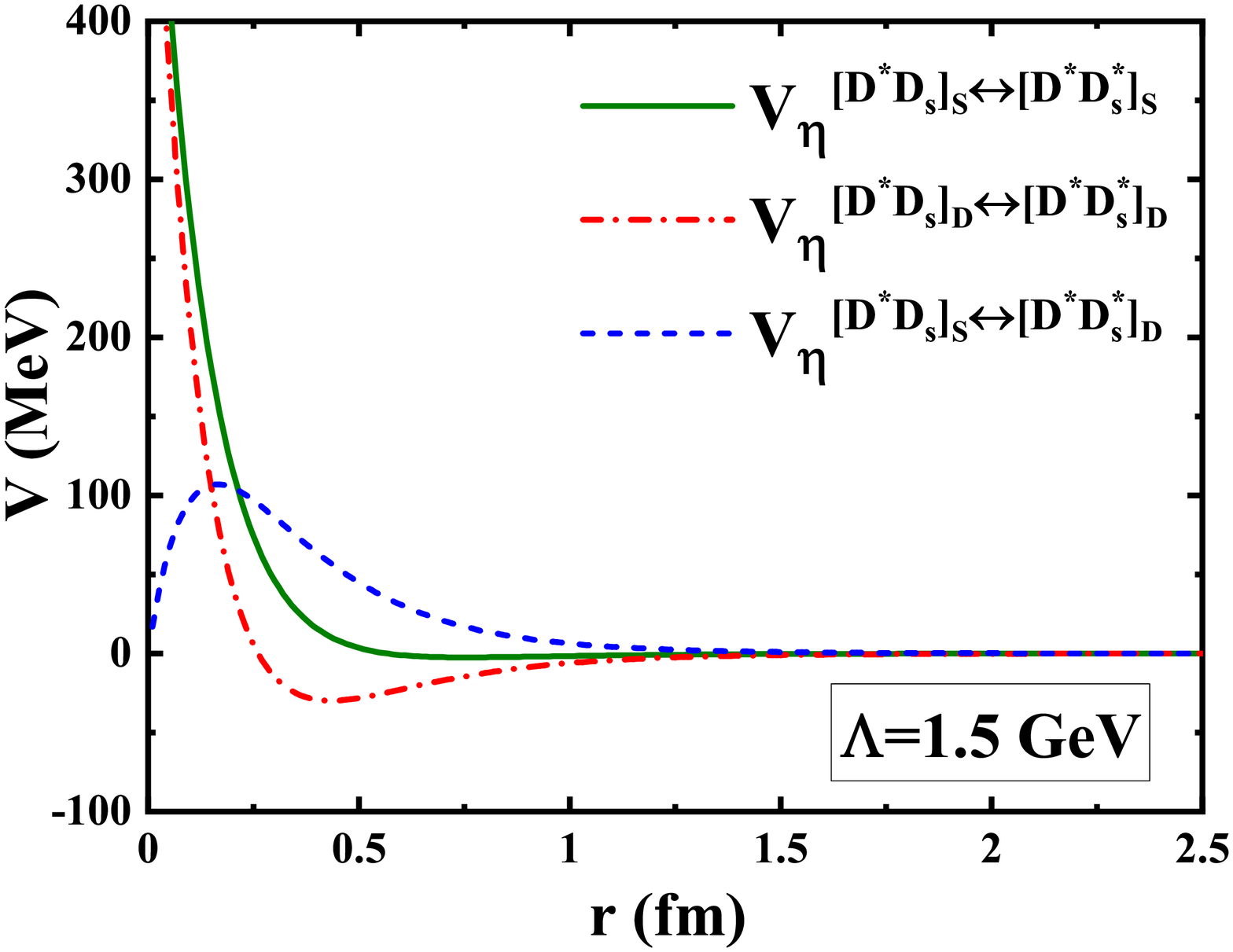}
\includegraphics[width=2.3in]{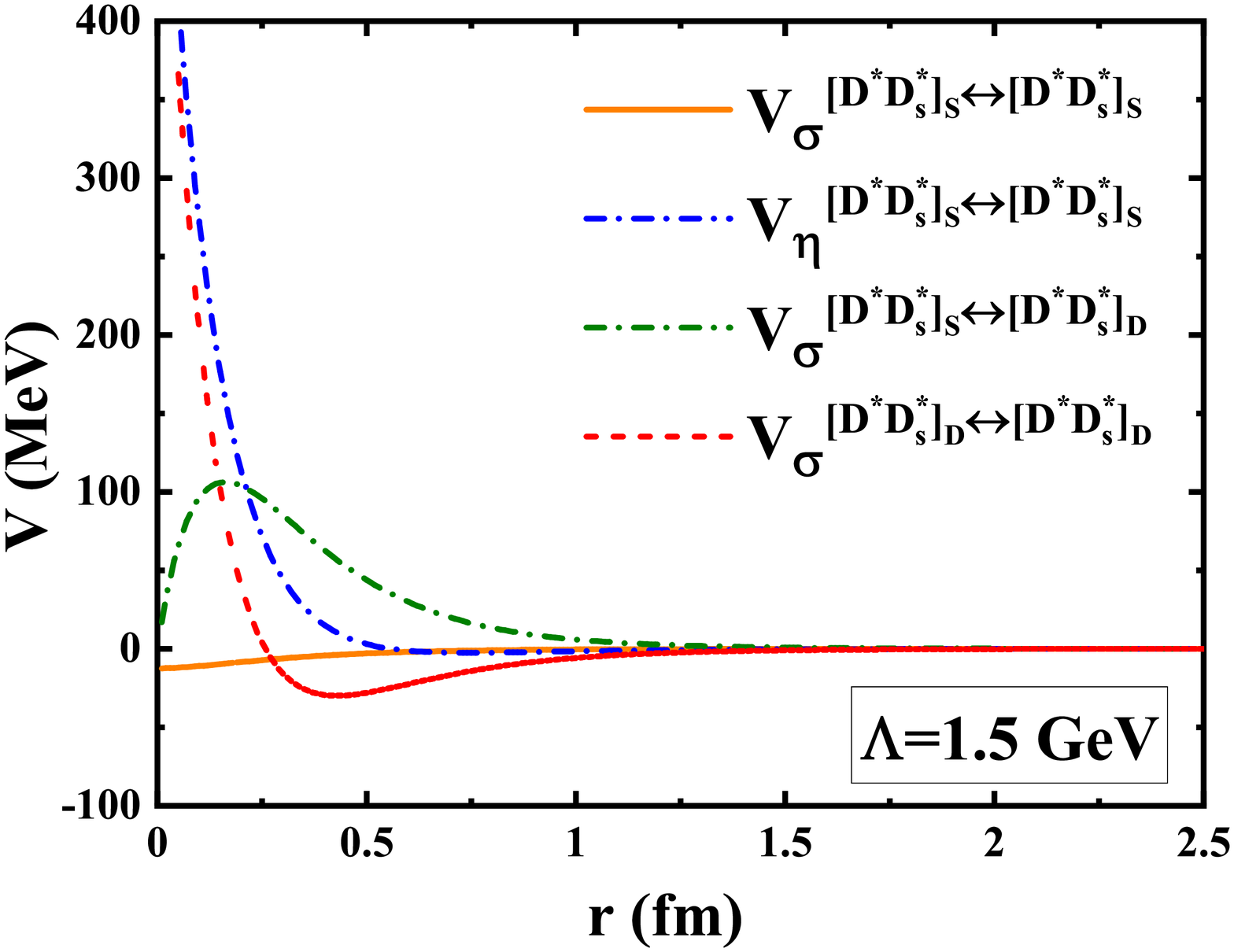}
\caption{(color online) OBE effective potentials for the $D^{*0}{D}_s^{-}\to D^{0}D_s^{*-}$, $D^{*0}{D}_s^{-}(D^{0}D_s^{*-})\to D^{*0}D_s^{*-}$, and $D^{*0}D_s^{*-}\to D^{*0}D_s^{*-}$ processes with $\Lambda=1.5$ GeV.}\label{sdwave12}
\end{figure*}

In our calculation, we take the same range of the cutoff value $1.0\leq\Lambda\leq2.0$ GeV to produce the phase shifts for the coupled $D^{*0}{D}_s^{-}/D^{0}D_s^{*-}/D^{*0}D_s^{*-}$ systems. 
As is well known, when the phase shift satisfies $\delta=(n+1/2)\pi$ with $n=0,1,2,\ldots$, the cross section $\sigma(E)$ takes the maximum $\sigma(E_0)$. It corresponds to a resonance, its mass and width are
\begin{eqnarray}
M=E_0,\quad\quad\quad    \Gamma=2/\left(\frac{d\delta(E)}{{d}E}\right)_{E_0},
\end{eqnarray}
respectively.

Finally, we cannot find any possible resonances for the $S$ wave single $D^{*0}{D}_s^{-}$ and $D^{0}D_s^{*-}$ when cutoff $\Lambda$ is taken from 1.0 to 2.0 GeV. If the $S-D$ wave mixing is considered, unfortunately, we cannot find any resonant evidences for the single $D^{*0}{D}_s^{-}$ and $D^{0}D_s^{*-}$ systems either. Therefore, there doesn't exist possible $D^{*0}{D}_s^{-}$ and $D^{0}D_s^{*-}$ resonance candidates from the single channel analysis. Here, we can conclude that the newly $Z_{cs}$ is excluded as a $D^{*0}{D}_s^{-}$ or $D^{0}D_s^{*-}$ shape-type resonance.

In the following, we further consider the coupled channel effect to check whether the newly $Z_{cs}(3985)$ can be a possible Feshbach-type resonance from the $D^{*0}{D}_s^{-}/D^{0}D_s^{*-}/D^{*0}D_s^{*-}$ interactions. In Fig. \ref{sdwave12}, we present the OBE effective potentials for the $D^{*0}{D}_s^{-}\to D^{0}D_s^{*-}$, $D^{*0}{D}_s^{-}(D^{0}D_s^{*-})\to D^{*0}D_s^{*-}$, and $D^{*0}D_s^{*-}\to D^{*0}D_s^{*-}$ processes with $\Lambda=1.5$ GeV. Here, the OBE effective potentials from $D^{*0}{D}_s^{-}\to D^{*0}D_s^{*-}$ scattering process has the same numerical values with those in the $D^{0}D_s^{*-}\to D^{*0}D_s^{*-}$ process. Compared to the $\sigma$ exchange, the $\eta$ exchange interactions are much stronger, which may play an important role in the discussed coupled systems. 

\begin{figure*}[!htbp]
\center
\includegraphics[width=2.3in]{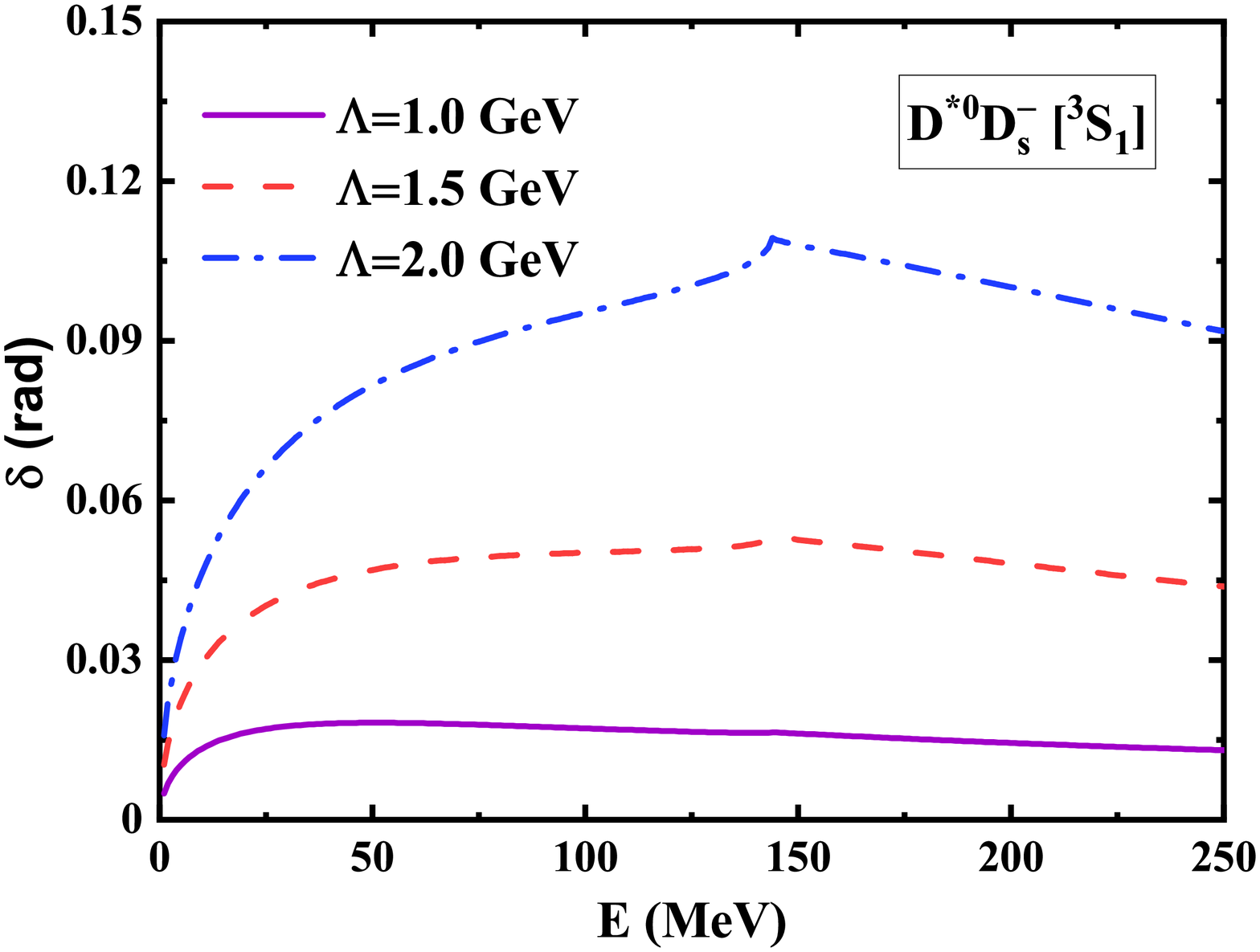}
\includegraphics[width=2.3in]{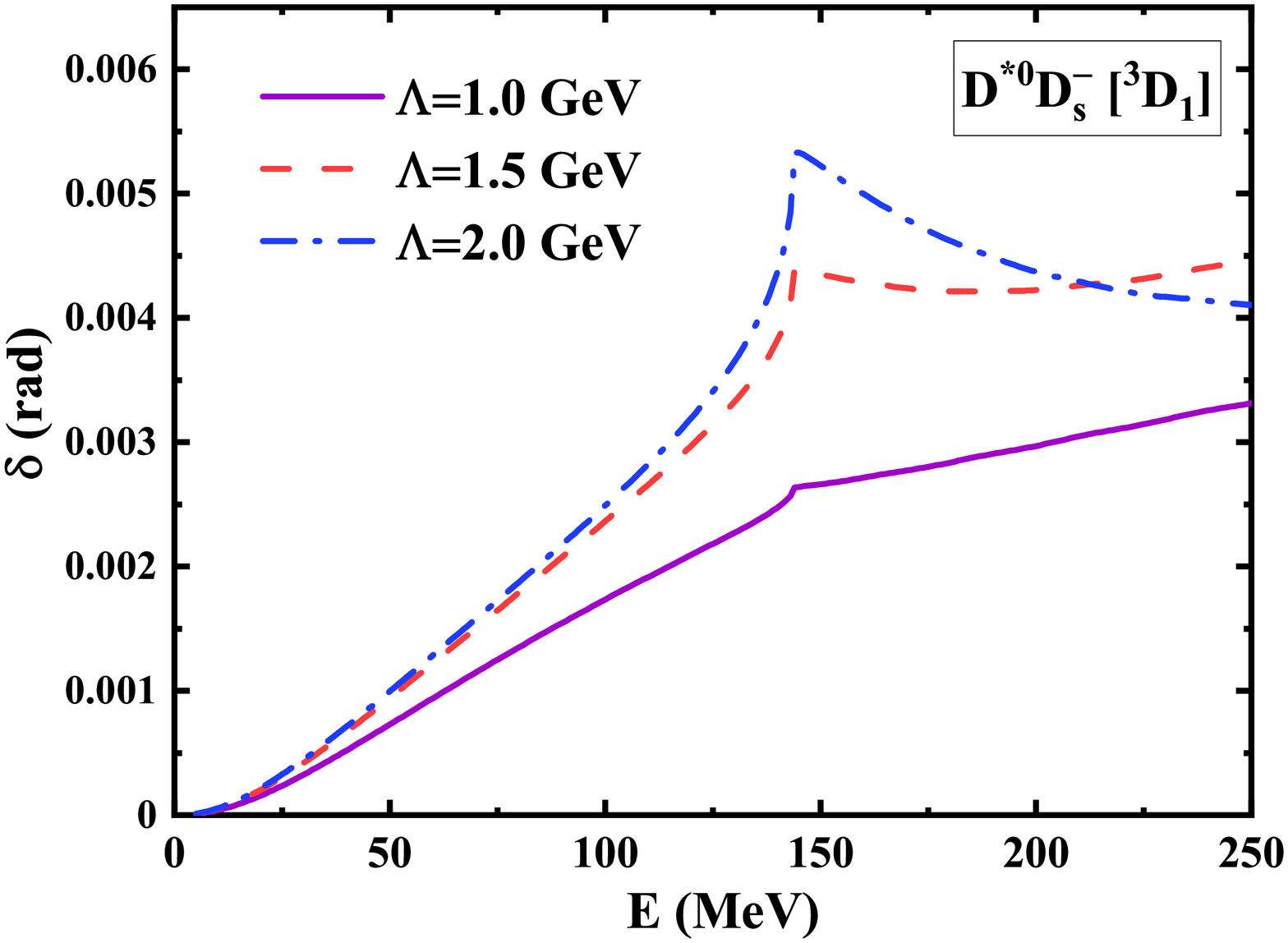}
\includegraphics[width=2.3in]{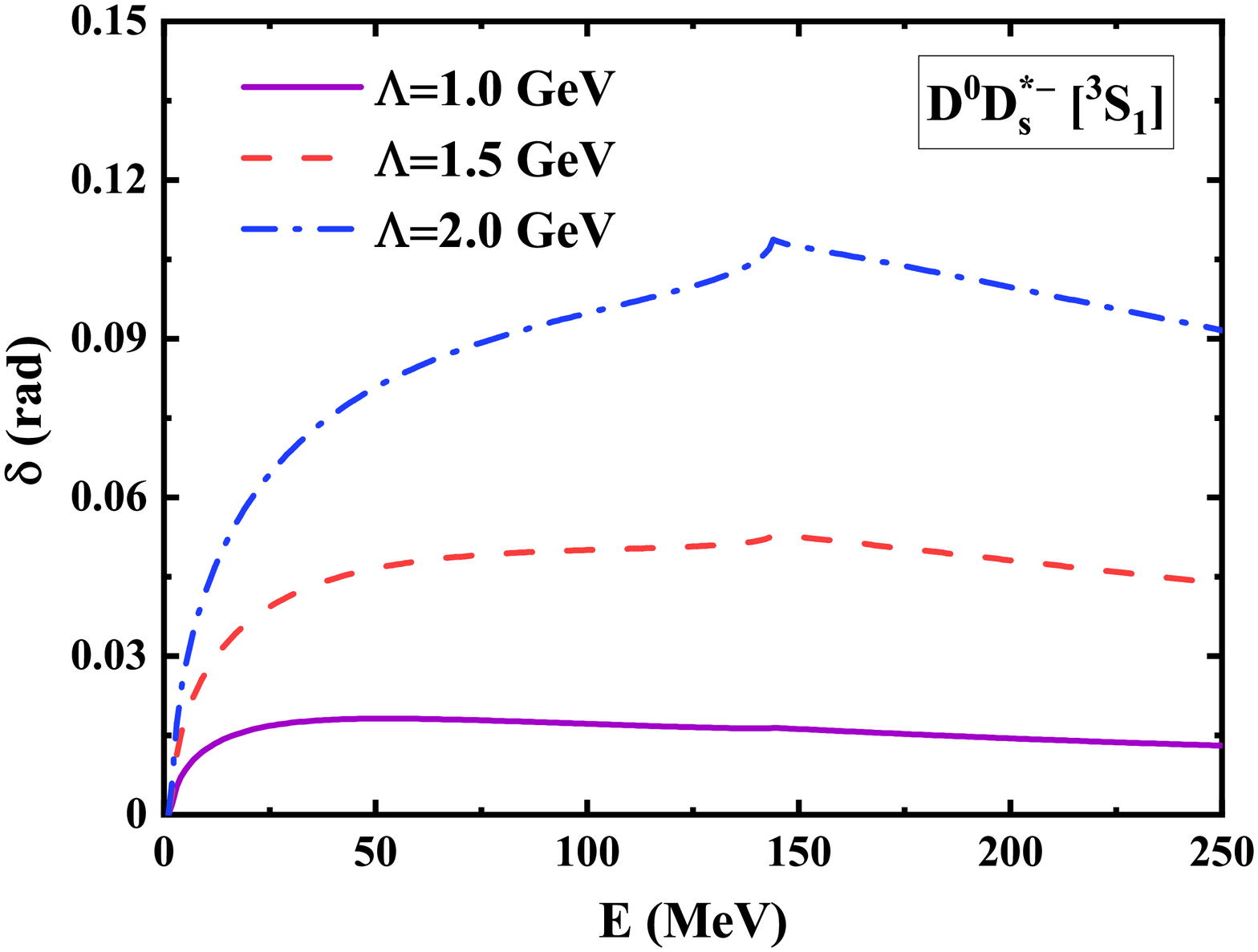}\\
\includegraphics[width=2.3in]{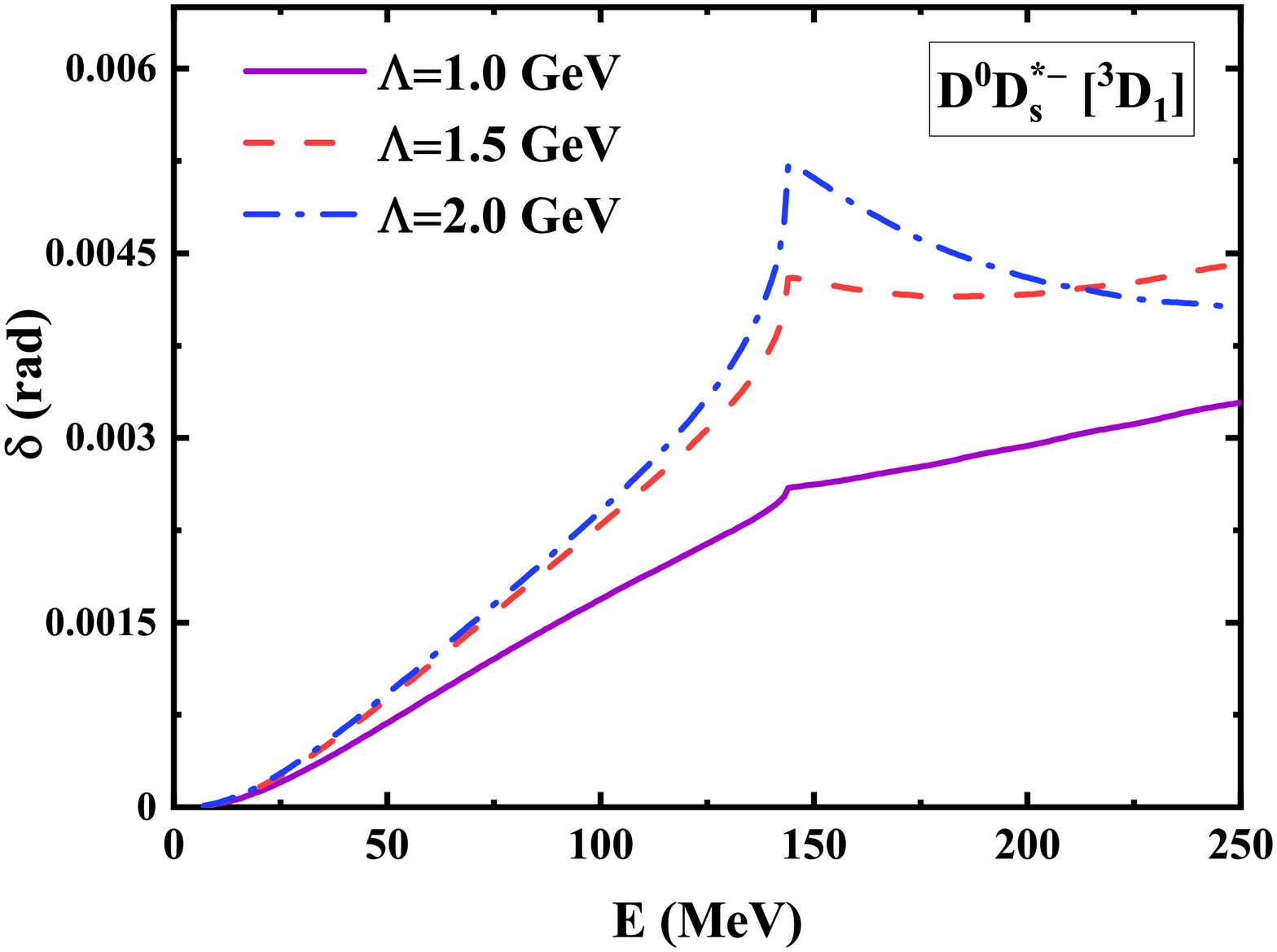}
\includegraphics[width=2.3in]{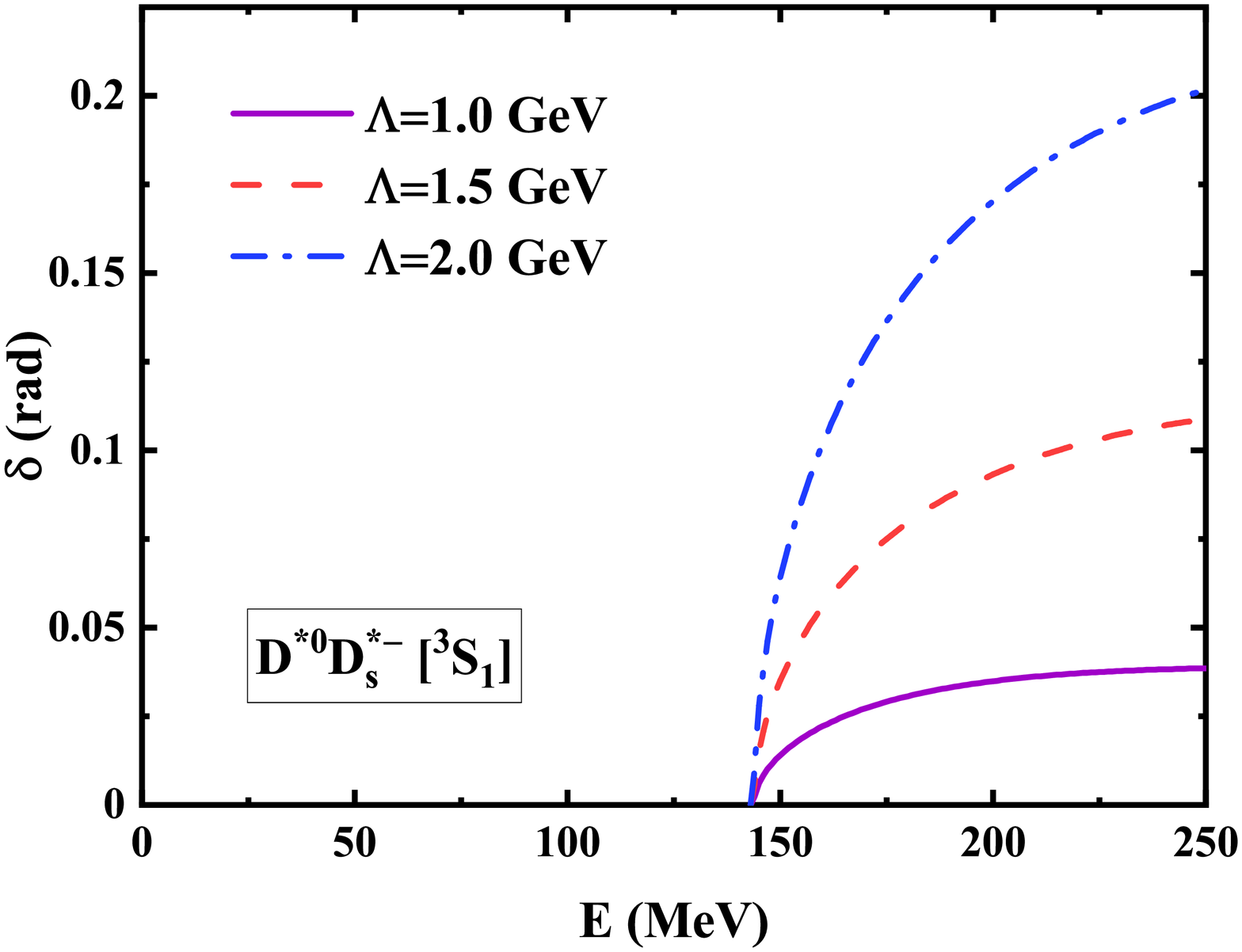}
\includegraphics[width=2.3in]{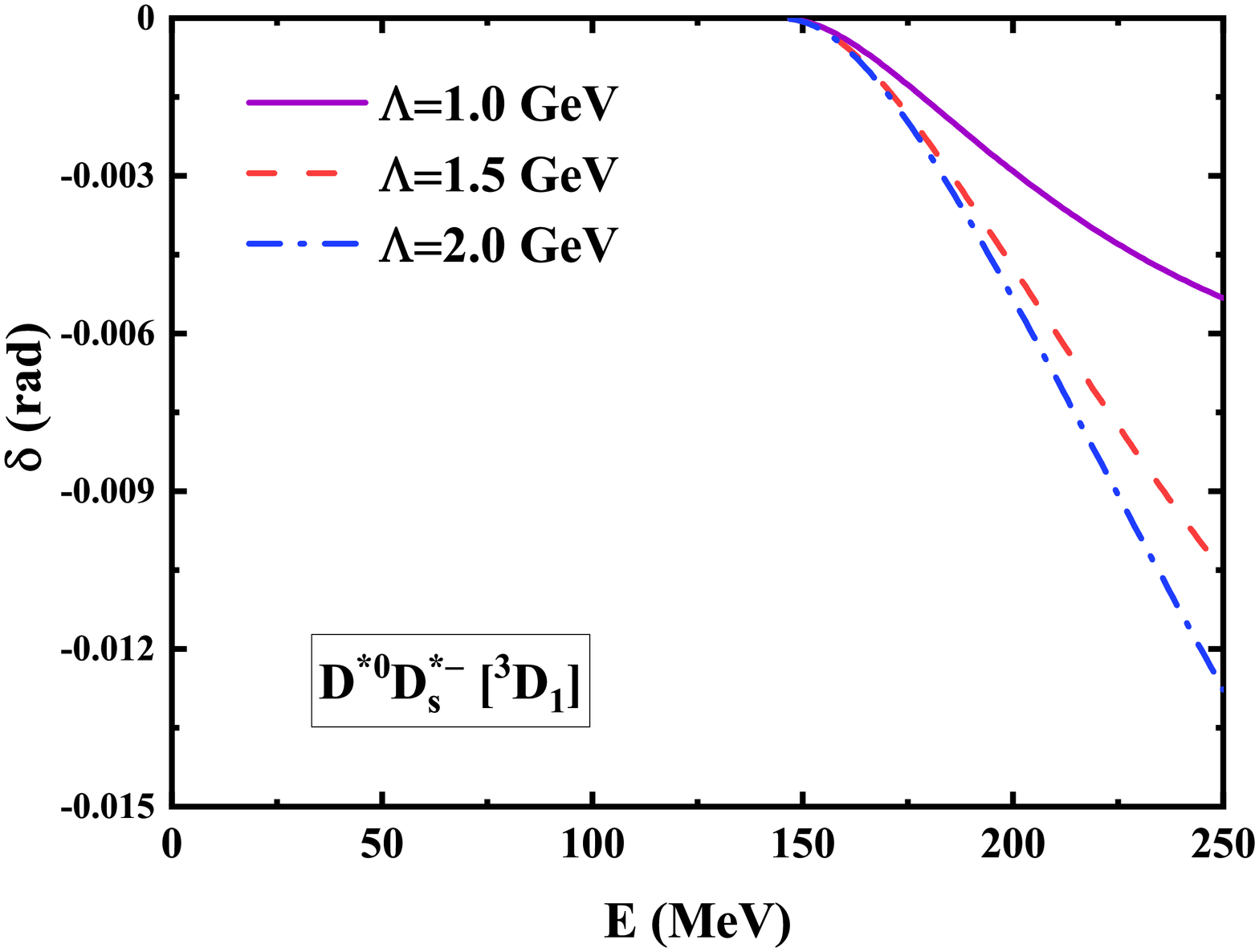}
\caption{(color online) Phase shifts for the $D^{*0}{D}_s^{-}-D^{0}D_s^{*-}-D^{*0}D_s^{*-}$ coupled systems with $\Lambda=1.0, 1.5$, and $2.0$ GeV.}\label{phase}
\end{figure*}

In Fig. \ref{phase}, we take three groups of typical cutoff values $\Lambda=1.0, 1.5, 2.0$ GeV to plot the energy $E$ dependence of the phase shift $\delta$ for the coupled $D^{*0}{D}_s^{-}/D^{0}D_s^{*-}/D^{*0}D_s^{*-}$ systems with $I(J^P)=1/2(1^+)$. As shown in Fig. \ref{phase}, with the increasing of the cutoff $\Lambda$, the OBE interactions become much stronger, which leads to the absolute values of the corresponding phase shifts for all channels become much larger. Also, small peaks appear at the $D^{*0}D_{s}^{*-}$ threshold in the phase shifts for all the $D^{*0}D_s^-$ and $D^0D_s^{*-}$ channels, which are interpreted as cusps. Most importantly, in the cutoff range of $1.0\leq\Lambda\leq2.0$ GeV, phase shifts for all discussed channels except the $D^{*0}D_s^{*-}|{}^3D_1\rangle$ rise up, but all their values are less than $\pi/2$, which indicates that no resonance exists. Thus, we can further conclude that the newly $Z_{cs}(3985)$ cannot be interpreted as a $D^{*0}{D}_s^{-}/D^{0}D_s^{*-}/D^{*0}D_s^{*-}$ Fechbash-type resonance.

So, according to our study, we can exclude the newly $Z_{cs}(3985)$ as the $I(J^P)=1/2(1^+)$ $D^{*0}{D}_s^{-}/D^{0}D_s^{*-}/D^{*0}D_s^{*-}$ shape or Fechbash resonance in the OBE model.


Apart from $S-D$ wave mixing, we also test if the newly $Z_{cs}$ can be a $P$ wave $D^{(*)}\bar{D}_s^{(*)}$ resonance. In this situation, the spin-orbit wave functions are
\begin{eqnarray*}
&&J^P=0^-: \,D^{*0}D_s^-|{}^3P_0\rangle,\, D^{0}D_s^{*-}|{}^3P_0\rangle,\, D^{*0}D_s^{*-}|{}^3P_0\rangle, \\
&&J^P=1^-:\, D^{*0}D_s^-|{}^3P_1\rangle,\, D^{0}D_s^{*-}|{}^3P_1\rangle,\, D^{*0}D_s^{*-}|{}^1P_1,{}^3P_1,{}^5P_1\rangle,\\
&&J^P=2^-:\, D^{*0}D_s^-|{}^3P_2\rangle,\, D^{0}D_s^{*-}|{}^3P_2\rangle,\, D^{*0}D_s^{*-}|{}^3P_2,{}^5P_2\rangle,
\end{eqnarray*}
and the corresponding matrix elements for the spin-spin interaction and tensor force operators are summarized in Eq. (\ref{matrix1})-(\ref{matrix2}). For the $P$ wave case, when we vary the cutoff value from 1.0 to 2.0 GeV, we still don't find any resonances. Therefore, we further find that the newly $Z_{cs}$ cannot be a $D^{(*)}\bar{D}_s^{(*)}$ resonance with $0^-$, $1^-$, or $2^-$.

\begin{eqnarray}
&&\left\{\begin{array}{c}
\bm{\epsilon}_1\cdot\bm{\epsilon}_3^{\dag}\\
\bm{\epsilon}_2\cdot\bm{\epsilon}_4^{\dag}\\
\bm{\epsilon}_1\cdot\bm{\epsilon}_3^{\dag}\bm{\epsilon}_2\cdot\bm{\epsilon}_4^{\dag}\end{array}
\right\} \to \left(\mathcal{I}\right)_{J^P=0^-,1^-,2^-},\label{matrix1}\\
&&\frac{i}{\sqrt{2}}(\bm{\epsilon}_2\times\bm{\epsilon}_3^{\dag})\cdot\bm{\epsilon}_4^{\dag}
  \to \left\{(1)_{0^-},   (0, 1, 0)_{1^-},    (1, 0)_{2^-}\right\},\nonumber\\
&&(\bm{\epsilon}_1\times\bm{\epsilon}_3^{\dag})\cdot(\bm{\epsilon}_2\times\bm{\epsilon}_4^{\dag})
   \to\nonumber\\
&&\left\{(1)_{0^-}, (\text{diag}(2,1,-1)_{1^-}, (\text{diag}(1,0)_{2^-}\right\},\nonumber\\
&&S(\hat{r},\bm{\epsilon}_2,\bm{\epsilon}_4^{\dag}) \to
    \left\{(2)_{0^-}, (-1)_{1^-},  (\frac{1}{5})_{2^-}\right\},\nonumber\\
&&S(\hat{r},i\bm{\epsilon}_2\times\bm{\epsilon}_3^{\dag},\bm{\epsilon}_4^{\dag})
    \to \left\{(\frac{-1}{\sqrt{2}})_{0^-},   (0, \frac{1}{\sqrt{2}}, 0)_{1^-},    (\frac{-1}{\sqrt{10}}, 0)_{2^-}\right\},\nonumber\\
&&S(\hat{r},\bm{\epsilon}_1\times\bm{\epsilon}_3^{\dag},\bm{\epsilon}_2\times\bm{\epsilon}_4^{\dag})
   \to\nonumber\\
&&\left\{(2)_{0^-}, \,
   \left(\begin{array}{ccc} 0 & 0 & -\frac{2}{\sqrt{5}} \\ 0 & -1 & 0 \\ -\frac{2}{\sqrt{5}} & 0 & \frac{7}{5} \\\end{array}\right)_{1^-},\,
\left(\begin{array}{cc} \frac{1}{5} & 0 \\ 0 & -\frac{7}{5} \\\end{array}\right)_{2^-}\right\}.\label{matrix2}
\end{eqnarray}

\section{Summary}\label{sec4}

In recent decades, people pay more and more attention on the study of multiquark states. In particular, the observations of the charged charmonium-like structures such as $Z_c(3900)^{\pm}$, $Z_c{\pm}(4020)$ and $Z_c^+(4430)$ provide important evidences of the existences of tetraquark state in the charm sector. Very recently, the BESIII Collaboration reported a new charged charmonium-like state $Z_{cs}^-(3985)$ in in the $K^+$ recoil-mass spectrum of the $e^+e^-\to (D^{*0}D_s^-/D^0D_s^{*-})K^+$ processes, which could be the first candidate of the charged hidden-charm tetraquark state with strangeness. 

Stimulated by the observation of the strange charmonium-like state $Z_{cs}(3985)$ and its behavior of being closed to threshold, in this work, we perform a systematic investigation on the $D^{*0}{D}_s^{-}/D^{0}D_s^{*-}/D^{*0}D_s^{*-}$ interactions by using the OBE model. In this situation, only the intermediate range interaction from the $\sigma$ and $\eta$ exchanges contributes to the coupled $D^{*0}{D}_s^{-}/D^{0}D_s^{*-}/D^{*0}D_s^{*-}$ system, and the $\eta$ exchange effective potential is dominant. These are remarkable different with the $D^{(*)}\bar{D}^{{*}}$ interactions since where the long range force from the $\pi$ exchange and the short range force from the vector meson $\rho/\omega$ exchanges are allowed, in particular, the $\pi$ exchange are the most important.

After gradually introducing the $S-D$ wave mixing and the coupled channel effect to study the phase shifts of the coupled $D^{*0}{D}_s^{-}/D^{0}D_s^{*-}/D^{*0}D_s^{*-}$ systems, finally, our results exclude the newly $Z_{cs}(3985)$ as a shape-type or Feshbach-type strange hidden-charm tetraquark resonance with $I(J^P)=1/2(1^+)$. In addition, we further study the $P-$wave $D^{*0}{D}_s^{-}/D^{0}D_s^{*-}/D^{*0}D_s^{*-}$ interactions, and find that the newly $Z_{cs}$ cannot be interpreted as a $D^{(*)0}D_s^{(*)-}$ resonance with $I(J^P)=1/2(0^-, 1^-, 2^-)$.

\section*{ACKNOWLEDGMENTS}

Rui Chen is very grateful to Yasuhiro Yamaguchi for his guidance in the numerical calculation and Xiang Liu for his helpful discussions. R. C. is supported by the National Postdoctoral Program for Innovative Talent.

\end{document}